\documentclass[sigconf, nonacm]{acmart}

\usepackage{graphicx}
\usepackage{xcolor}
\usepackage{booktabs}
\usepackage{enumitem}
\usepackage{multirow}
\usepackage[bottom]{footmisc}
\usepackage[frozencache=true,cachedir=minted-cache]{minted} 
\usepackage{url}
\usepackage{multicol}
\usepackage{tabularx}
\usepackage[normalem]{ulem}
\newcolumntype{Y}{>{\centering\arraybackslash}X}

\begin{document}

\newcommand{\fix}[1]{\textcolor{red}{#1}}
\newcommand{\added}[1]{\textcolor{blue}{#1}}

\newcommand{\cooltitle}[1]{\vspace{1mm}\noindent\textbf{#1}}

\newcommand{\idecode}[0]{non-notebook code}

\newcommand{\hist}[1]{(\includegraphics[height=2.5mm]{#1})}
\newcommand{\histt}[1]{\includegraphics[height=2.5mm]{#1}}
\newcommand{\enrique}[1] {\textcolor{blue}{\textbf{[Enrique: #1]}}}
\newcommand{\neil}[1] {\textcolor{purple}{\textbf{[Neil: #1]}}}
\newcommand{\derek}[1] {\textcolor{gray}{\textbf{[Derek: #1]}}}
\newcommand{\peggy}[1] {\textcolor{violet}{\textbf{[Peggy: #1]}}}
\newcommand{\cassie}[1] {\textcolor{orange}{\textbf{[Cassie: #1]}}}

\title{Error Identification Strategies for Python Jupyter Notebooks}

\author{Derek Robinson}
    \affiliation{
        \institution{Department of Computer Science \\University of Victoria}
        \city{Victoria}
        \country{Canada}}
    \email{drobinson@uvic.ca}

\author{Neil A. Ernst}
    \affiliation{
        \institution{Department of Computer Science \\University of Victoria}
        \city{Victoria}
        \country{Canada}}
    \email{nernst@uvic.ca}

\author{Enrique Larios Vargas}
    \affiliation{
        \institution{Department of Computer Science \\University of Victoria}
        \city{Victoria}
        \country{Canada}}
    \email{elariosvargas@uvic.ca}

\author{Margaret-Anne D. Storey}
    \affiliation{
        \institution{Department of Computer Science \\University of Victoria}
        \city{Victoria}
        \country{Canada}}
    \email{mstorey@uvic.ca}

\renewcommand{\shortauthors}{D. Robinson, N. Ernst, E. Larios Vargas, M.-A. Storey}

\begin{abstract}
Computational notebooks---such as Jupyter or Colab---combine text and data analysis code.
They have become ubiquitous in the world of data science and exploratory data analysis. 
Since these notebooks present a different programming paradigm than conventional IDE-driven programming, it is plausible that debugging in computational notebooks might also be different. 
More specifically, since creating notebooks blends domain knowledge, statistical analysis, and programming, the ways in which notebook users find and fix errors in these different forms might be different.
In this paper, we present an exploratory, observational study on how Python Jupyter notebook users find and understand potential errors in notebooks. 
Through a conceptual replication of study design investigating the error identification strategies of R notebook users, we presented users with Python Jupyter notebooks pre-populated with common notebook errors---errors rooted in either the statistical data analysis, the knowledge of domain concepts, or in the programming. 
We then analyzed the strategies our study participants used to find these errors and determined how successful each strategy was at identifying errors.
Our findings indicate that while the notebook programming environment is different from the environments used for traditional programming, debugging strategies remain quite similar.
It is our hope that the insights presented in this paper will help both notebook tool designers and educators make changes to improve how data scientists discover errors more easily in the notebooks they write.
\end{abstract}

\maketitle

 \section{Introduction}

Jupyter Notebook\footnote{https://jupyter.org/} is an open-source, browser-based programming environment that allows users to weave rich text, code, equations, and visualizations into a single human-readable document. 
Jupyter Notebooks and other computational notebooks, such as Google Colab\footnote{https://colab.research.google.com/}, RMarkdown Notebooks\footnote{https://rmarkdown.rstudio.com/}, and Azure Notebooks\footnote{https://notebooks.azure.com/}, have become immensely popular for anyone who wishes to perform data analysis or exploration tasks. 
The Jupyter platform specifically has grown exponentially since 2015, with over 6 million publicly available Jupyter Notebooks currently residing on GitHub alone \cite{Parente2014}.
Despite its popularity, Jupyter Notebook users have mixed opinions about the process of debugging~\cite{Chattopadhyay2020}. 
Some praise its ability to help them find errors quickly, but others complain about a poor debugging experience~\cite{Chattopadhyay2020}. 
This has given us some insight into how users \textit{feel} about debugging Jupyter Notebooks, but not much about their actual debugging processes.

Debugging involves two phases. 
The first phase involves identifying what and where the error is, and the second phase involves determining how best to fix the error \cite{Zeller2009programs}.
We have many insights about the process of debugging software systems and programs in general~\cite{Alaboudi2021,Whalley2021,Murphy2008}, but we lack similar insights for computational notebooks.
To the best of our knowledge, there are a few studies which examine how computational notebooks are debugged. For instance, Yang \emph{et al.} \cite{yang21ase}, focused on a tool for improving code understanding with program synthesis but did not synthesize strategies used in debugging. 
Without knowing \emph{how} computational notebooks are debugged, it is difficult to support---whether with tools or processes---the debugging of computational notebooks.

Motivated by the sheer popularity \cite{Parente2014} of the Jupyter Notebook platform across many disciplines, we aimed to identify strategies adopted by users of Jupyter Notebooks in the first phase of debugging, i.e., strategies for \emph{identifying} errors in data analysis notebooks.  
Understanding debugging strategies may provide educators and students with valuable knowledge about how errors are found in computational notebooks, improving how they teach or learn error-finding methods. 
Although our findings are not aimed at any specific community of Jupyter Notebook users, as the platform is used across many different domains \cite{Cardoso2018, Smith2016, Weiss2020}, we hope that our results may serve as recommendations for users of Jupyter Notebooks who may not have a strong background in the process of debugging.

Our work focused on this research question:\\
\indent{\textbf{What strategies do data scientists use to find statistical, data, and programming errors in Python Jupyter Notebooks?}}

Our study is a conceptual replication of 
a work-in-progress study of how experts and novices debug RMarkdown documents \cite{mcnamara22}. All source materials \cite{bodwin22zenodo} were provided by the authors of the R study. We replicated the R study design and
performed an observational study with 14 participants tasked with finding and understanding errors in one of four Jupyter Notebooks. We translated the RMarkdown study materials \cite{bodwin22zenodo} into Python, retaining the same pre-populated errors in the statistics, data, and programming code and the same supporting analysis text as designed in the original study  materials~\cite{mcnamara22}. The participants could use any error-finding method they chose and were not time-constrained. 
We observed that participants followed seven different strategies.
We characterized the strategies according to the frequency of use and their success (which are not necessarily the same!).
Consulting external resources via a search engine was the most \textit{common} strategy. 
However, the most \textit{successful} strategy was Expectation Confirmation, where there was a mismatch between what explanatory markdown cells claimed and what the code actually did. 
On average, our participants found approximately 40\% of the errors in the notebook they analyzed. 

\section{Background}

Jupyter Notebooks are the ``de-facto standard'' for data scientists~\cite{Perkel2018}, and much has been learned about how reproducible they are, the quality of the code written in them, and the narratives that describe the analyses within them \cite{Pimentel2019, Wang2020Better, Wang2020Restoring, Rule2018}. These studies agree that notebook code is frequently low-quality and error-prone. 
Closely related to our work is that from Yang \emph{et al.} \cite{yang21ase}. They report on a tool to support bug detection in Kaggle notebooks, which they characterized as `data wrangling code'. Like us, they show this style of development is quite different than pure source code approaches and prone to errors, yet not well supported by tools. They introduce WrangleDoc, a program synthesis technique to summarize code in order to facilitate debugging. We did not examine such summarization approaches in our study, and our Jupyter instance contained no plugins. 

While the work of Yang \emph{et al.} highlights the potential problems with data science code, there is still much we do not understand. In particular, their study looked at specific tool support using a documentation approach. We focused instead on the strategies leading to the discovery of notebook errors, the starting point of the debugging process.

\cooltitle{Challenges of using notebooks.} While the literature on notebook error detection is limited, other research has revealed the challenges of using notebooks. 
Chattopadhyay \emph{et al.} identified the main pain points of using computational notebooks, including Jupyter \cite{Chattopadhyay2020}. 
Of interest to this study is the \emph{Manage Code} pain point which mentions that without sufficient software engineering support, debugging, writing code, managing dependencies, and testing relies on \emph{ad-hoc} workarounds.
Specifically, they found that writing code in notebooks efficiently requires knowledge of all the function names and classes, plus the use of a second window to search for online resources such as documentation~\cite{Chattopadhyay2020}. 
They also observed a divide in how their participants felt about debugging in notebooks. 
Some participants were able to find errors in a notebook quickly, but others found that debugging was a horrible experience when they had to rely on \mintinline{Python}{print()} statements.
In addition, they found that testing in computational notebooks was difficult as there is no standard method to test a notebook. 
Some study participants wrote test cases in the same notebook, while others created a new notebook for testing.

\cooltitle{Debugging traditional programs.} There is plenty of related research on debugging conventional programs. We mention some closely related studies here.
Murphy \emph{et al.} present a qualitative study of the debugging strategies employed by computer science students \cite{Murphy2008}. 
They observed three distinct categories of strategy: the good, the bad, and the quirky. 
The good strategies (or effective strategies) included \emph{gaining domain knowledge, tracing, testing, understanding the code, using resources, using tools, isolating the problem, pattern matching, and considering alternatives}. 
They also identified that many students employed strategies that were less effective (the bad).
These ``bad'' strategies were the same as the effective strategies, but employed less effectively. 
Finally, the quirky strategies were ones which surprised Murphy \emph{et al.}.

\cooltitle{Hypothesis-driven debugging.} Alaboudi and Latoza authored two papers that relate to our study. 
The first paper, titled ``Using Hypothesis as a Debugging Aid'' \cite{Alaboudi2020}, describes two studies.
In the first study, they observed live-stream videos of developers' programming activities. 
Their second study was a controlled study of 25 participants tasked with debugging three API misuse problems. 
Overall, they observed that developers found it challenging to formulate a reasonable hypothesis about a potential error. 
In the second paper, ``An Exploratory Study of Debugging Episodes'', Alaboudi and Latoza observed 15 live-streamed programming sessions (in C, C\#, JavaScript) \cite{Alaboudi2021}. 
They found that developers spent 48\% of their programming sessions debugging. 
They also found that no single activity dominated a debugging session, with developers spending varied amounts of time on different activities. 
Additionally, they observed significant differences between long and short debugging episodes. 
Short debugging episodes focused on editing and testing code, while long debugging episodes involved various activities, such as consulting external resources, and inspecting program state in addition to testing and editing.

\cooltitle{Student approaches to debugging.} Like our study, Whalley \emph{et al.} examined students' thoughts about their debugging process (in \idecode{}). They examined whether reflecting on the debugging process helps students perceive a need for change in their approach, and if they perceive value in a structured, formal debugging process \cite{Whalley2021}. 
Whalley \emph{et al.} used semi-structured interviews to answer their research questions. 
Their analysis uncovered themes about code comprehension, bug location, information gathering strategies, challenges locating bugs, emotions felt during the debugging process, and the value students give to a formal debugging process. 
When students were asked to reflect on their debugging process, their comments referred to both high- and low-level activities. 
High-level activities included activities such as reading code, search space reduction, and hypothesis forming.
Most of the reflections shared about debugging were about low-level activities, such as where to place print statements, code tracing, and examining function parameters and return values. 
Many students perceived debugging as inefficient, likely due to the lack of a formal process to follow. 
One-third of the participants described their debugging process as flawed, and they universally described their hypothesis-forming method as imprecise, opting to guess and check instead.

\cooltitle{Data science debugging.}
Debugging traditional programs is well studied and the works discussed above are a very small subset of the work available on debugging. 
In contrast to the debugging of traditional programs, data science work can be quite different \cite{wong19}. For example, it is a common part of the workflow for scientists to re-run analyses (e.g., as part of exploratory data analysis).
This might happen when, for example, removing outliers or experimenting with different hyper-parameters. 
Thus, research has looked at supporting such experimental workflows in order to manage versions of notebooks~\cite{Weinman2021,Kery2018}, and support cleanup and refactoring \cite{Head2019}. 
Related to that is work that uses tools to debug data flows in large (non-notebook) data analytics pipelines~\cite{Rezig2020}, or support  statistical transparency with multiverse analysis \cite{Dragicevic2019}. None of this work focused on \emph{how} errors were detected. 
However, Brown \emph{et al.} introduces four error types present in data analysis \cite{Brown2018}.

\section{Materials and Methods} \label{sec:materials-and-methods}
To answer our research question, we observed the behaviour of 14 participants as they browsed and debugged existing Jupyter Notebooks that contained errors. 
The observations took place over Zoom, and participants shared their screens. 
We recorded video and audio of the meetings for a later qualitative analysis of the strategies our participants used. 
The data collection was conducted from November 2020 to January 2021.

\subsection{Participants}

Table \ref{tbl:participants} summarizes participant demographics. 
A majority (8) of the participants were from the domain of computer science or computer science combined with either music, biochemistry, or life sciences. 
The other (5) participants were each students in one of chemistry, physics, civil engineering, software engineering, and electrical/computer engineering. 
The remaining participant was a professional who worked in the education domain. 
Nine participants had used computational notebooks for less than one year, and the remaining five used them for between one and three years.
All participants stated that Jupyter Notebooks was their computational notebook of choice, with two also using Google Colab. 
\begin{table*}
    \small
    \centering
    \caption{Participant Demographics}
    \label{tbl:participants}
    \begin{tabular}{ccccc}
        \toprule
        Participant & Role & Domain & Notebook Experience (yrs)  \\
        \midrule
        P1 & Master's & Electrical \& Computer Engineering & < 1 \\ 
        P2 & Master's & Chemistry & 1-3  \\ 
        P3 & Undergraduate & Computer Science \& Biochemistry &  < 1 \\ 
        P4 & Undergraduate & Computer Science \& Life Sciences & 1-3 \\ 
        P5 & Master's & Computer Science &  < 1  \\ 
        P6 & Undergraduate & Physics & 1-3  \\ 
        P7 & Master's & Civil Engineering &  < 1  \\ 
        P8 & Undergraduate & Software Engineering &  < 1  \\ 
        P9 & Undergraduate & Computer Science & 1-3  \\ 
        P10 & Educational Specialist &  Education &  < 1 \\ 
        P11 & Undergraduate & Computer Science \& Music &  < 1 \\ 
        P12 & Undergraduate & Computer Science \& Music &  < 1 \\ 
        P13 & Doctoral & Computer Science & 1-3 \\ 
        P14 & Undergraduate & Computer Science \& Music &  < 1  \\
        \bottomrule
    \end{tabular}
\end{table*}

We recruited participants by contacting instructors of three 400/500-level courses with data science themes, by posting on online communities, and through personal contacts. 
Participation in our study was voluntary, but we encouraged participation by providing a \$25 Amazon gift card to the first 12 respondents.
Our study was approved by our institutional review board.

\subsection{Study Materials}
\label{sec:material}

The notebooks used for our study were translated from R notebooks created as part of an in-progress study to investigate data scientists' debugging behavior~\cite{mcnamara22}. The R notebooks were written by statistics and data science education researchers~\cite{bodwin22zenodo} and 
covered two different topics (NBA Player of the Week and the 2011 Spain Election). 
Each notebook had three versions: A, B, and C. 
Errors were introduced into versions A and B of the original R notebooks by two members of the R study \cite{mcnamara22} team, the C notebook had no errors.
Table \ref{tbl:participantpernotebook} lists the number of errors per notebook. 

We translated the R notebooks (including errors) to Python, and the translations were verified by a third party, experienced in Python, statistics, and data science.
Additionally, the first author of this paper verified that the Python translations returned the same data, visualizations, and values. The Python and R Notebooks are available at~\cite{bodwin22zenodo}, along with a complete list of the errors.

We retained the error classification system from the R Study, which identified embedded errors as: \textbf{data}, \textbf{statistical}, and \textbf{programming} \cite{bodwin22zenodo}. 
In the notebooks, error types were not mutually exclusive and a given error could be a \textbf{programming} error in addition to a \textbf{data} or \textbf{statistical} error. 
These three different types of errors align with the different categories of errors defined by Brown \emph{et al.}~\cite{Brown2018}. 
We describe the three categories of errors below and provide examples using the NBA Player of the Week notebook(s). 

\begin{listing}
    \centering
    \begin{minted}[fontsize=\small, frame=single, numbersep=-10pt, linenos, breaklines, breakbefore={.s}]{Python}
    nba = nba.assign(
        Height = pd.to_numeric(nba['Height'].str.replace('cm','').str.replace('-[0-9]*','')),
        Weight = pd.to_numeric(nba['Weight'].str.replace('kg','')))
    \end{minted}
    \caption{A data error: assumes all data is in cm/kg.}
    \label{lst:dataerror}
\end{listing}

\textbf{Data} errors occur when a notebook does not fully explore the dataset, or when the format of the data is misunderstood.
Listing~\ref{lst:dataerror} shows an example of a \textbf{data} error: the \mintinline{Python}{Height} column is of type string and is assumed to either contain the centimeter unit or a string value representing a measurement. 
Similarly, it is assumed that the measurements in the \mintinline{Python}{Weight} column are all in kilograms, with some measurements containing the kilogram units. 
In fact, the \mintinline{Python}{Height} column has units of either centimeters, such as 203cm, or feet-inches, such as 6-11. 
The \mintinline{Python}{Weight} column has measurements which are in kilograms and contain the kilogram units, or are measurements in pounds that contain no units.
The above code cell removes the centimeter unit by calling \mintinline{Python}{str.replace('cm','')}. The inches measurement is also removed by calling \mintinline[breaklines, breakbefore=(]{Python}{str.replace('-[0-9]*','')}. 
The same is done with the \mintinline{Python}{Weight} column, removing the kilogram units via \mintinline{Python}{str.replace('kg','')}. 
Thus, the column is incorrectly cleaned as the above code cell performs no unit conversions. 
This leaves the \mintinline{Python}{Height} and \mintinline{Python}{Weight} columns in mismatched units without any unit identifier.

\begin{listing}
    \centering
    \begin{minted}[fontsize=\small, frame=single, xleftmargin=\parindent, numbersep=-10pt,linenos, breaklines]{Python}
    my_test = ttest_ind(
        x1 = nba[(nba['Position'] == 'PG')]['Height'],
        x2 = nba[(nba['Position'] == 'SG')]['Height'],
        alternative = 'smaller')
    \end{minted}
    \caption{A statistical error: using a 1-sided t-test when a 2-sided t-test is the proper choice.}
    \label{lst:statserror}
\end{listing}

\textbf{Statistical} errors occur either when an incorrectly chosen statistical test or visualization is used, or when a correctly chosen test or visualization is wrongly interpreted by the user. 
In Listing~\ref{lst:statserror}, the goal is to determine if a statistical difference between the average height of point guards and shooting guards exists using a t-test. 
The error is that the \mintinline{Python}{alternative} parameter is set to \mintinline{Python}{smaller}, indicating a one-sided t-test. 
This parameter should be set to `two-sided' as the goal was to determine whether or not a statistical difference exists, rather than which average was smaller.

Lastly, \textbf{programming} errors occur when a code cell does not achieve the goal stated in the preceding markdown cell.
The goal of Listing~\ref{lst:programmingerror} is to filter the \mintinline{Python}{nba} dataframe so that it only contains unique players. 
While this code cell does output a set of unique players, a copy is returned, which is not saved to the \mintinline{Python}{nba} dataframe. 
Through the remainder of this notebook, the original \mintinline{Python}{nba} dataframe is used, and thus the code has an error and does not achieve its goal.

\begin{listing}
    \centering
    \begin{minted}[fontsize=\small, frame=single, xleftmargin=\parindent, numbersep=-10pt,linenos,breaklines, breakbefore = .]{Python}
    nba.groupby('Player').agg(
        Height=('Height', 'median'),
        Weight=('Weight', 'median'),
        Position=('Position', 'first'))
    \end{minted}
    \caption{A programming error: using the original, not the filtered data-frame.}
    \label{lst:programmingerror}
\end{listing}


\begin{table*}[!ht]
    \small
    \centering
    \caption{Number of Participants and Errors per Notebook\\(D: Data Error S: Statistical Error P: Programming Error)}
    \label{tbl:participantpernotebook}
    \begin{tabularx}{1.75\columnwidth}{lYYYYY}
     \toprule
     Notebook & \# of Participants & Participants & \# of Errors & Distribution of Errors & Size (\# of Code Cells)\\
     \midrule 
     nba\_analysis\_A & 4 & P1, P8, P10, P13& 6 & D:3 S:3 P:2 & 10\\ 
     nba\_analysis\_B & 3 & P2, P9, P12 & 4 & D:0 S:2 P:3 & 12\\ 
     elections\_analysis\_A & 4 & P3, P5, P6, P14 & 10 & D:2 S:7 P:5 & 12\\ 
     election\_analysis\_B & 3 & P4, P7, P11 & 10 & D:2 S:9 P:3 & 12\\
     \bottomrule 
    \end{tabularx}
\end{table*}

\subsection{Jupyter Notebook Study Design}
Each participant was tasked with finding potential errors in one of the four notebooks which contained errors (Versions A or B).
Version C was shown to them after their analysis if they wanted to see an error-free version. 
We aimed to balance the number of participants analyzing each notebook (see Table \ref{tbl:participantpernotebook}).  
This task was open-ended in that the participants were allowed to use any method they liked to find potential errors. 
The only specific instructions given were for them to think aloud whenever possible and to notify the researcher when they thought they had found an error.

We performed two rounds of pilots (with members of our research group) to improve the study task and to confirm our study would provide sufficient observations on error finding strategies.  
The feedback from the pilots helped us improve the study materials. 
The supplementary materials contain the task description, interview questions, and Jupyter Notebooks~\cite{bodwin22zenodo}. 

At the start of each study session, we described the task and emphasized that our aim was not to test their skills. 
Participants were then presented with a Jupyter Notebook and informed that any of the notebook components might contain errors that they should try to identify. 
We mentioned they could modify the notebook, search the documentation, or use the internet for help.

Each study session consisted of two phases: an observational phase and an interview phase. During the observational phase, participants analyzed the notebook for errors while one researcher observed their behaviours and took notes. Once the participant was satisfied with their analysis of the notebook, we held the interview phase of the study. 

The interview began with unstructured questions, using notes from our observations to guide our follow-up questions. Asking these questions immediately after the participant had performed their task was important as their strategies were still fresh in their mind. These unstructured questions were asked to gain insights into a specific approach and why it was used.  
Following the unstructured part of the interview, additional questions were asked about the participant's domain of study, how long they had been using Jupyter Notebooks, and the computational notebook they used most often. 
The complete list of these additional questions is available in a replication package \cite{bodwin22zenodo}.

\subsection{Data Collection and Analysis}

The Zoom video recordings of the studies were uploaded and we analyzed the recordings directly using ATLAS.ti 8\footnote{https://atlasti.com/product/what-is-atlas-ti/.}. 
We used an open coding process to code all activities performed by our participants. 
The first author of this paper performed the initial coding. 
After the initial coding cycle, discussion sessions were held with the second and fourth authors, where the codes were further analyzed and compared with the findings from previous participants, and refined in an iterative manner.

Throughout our discussion sessions, we identified emergent higher-level groups for the codes and merged some codes: 

\begin{itemize}
 	\item \textbf{Action:} An action that a participant performed.
 	\item \textbf{Docs:} A specific documentation website that a participant visited.
    \item \textbf{Online Resource:} An online resource other than documentation that a participant visited. 
    \item \textbf{Reasoning:} A reason for performing an action or a reason for why something was an error.  
    \item \textbf{Participant Attribute:} To describe a participant. 
\end{itemize}
 
Throughout our discussion sessions, we noticed many \emph{actions} were performed together. We called these connected sets of actions \emph{strategies}.
The first author analyzed the raw data again to identify and code strategies from each group of \emph{actions}. 

Once we identified strategies, we analyzed videos again to determine the success rate of each strategy. 
Whenever we observed a participant analyzing an erroneous cell, we entered their chosen strategy into a spreadsheet, along with the type of error they were working on and whether that strategy was successful or not. 

\section{Findings}

Our research question asked what strategies data scientists use to find statistics, data/domain, and programming errors.
Our analysis reveals (a) \textit{actions} and (b) \textit{strategies} that our participants employ to find errors in Python Jupyter notebooks. Additionally, we present (c) the \textit{relationship between strategies and error-finding success}. Tables~\ref{tbl:timeaction} and~\ref{tbl:strategytable} show the entire list of actions and strategies identified in our exploratory observational study. Finally, in Table~\ref{tbl:relatingstrats}, we present how strategies relate to the different error types. 
We describe each action, strategy, and their respective relationship with an error type below.   

\subsection{Actions Taken}

\begin{table*}
    \small
	\centering
	\caption{Actions Taken in Error Identification.}
	\label{tbl:timeaction}
	\begin{tabular}{lccc}
		\toprule
		Action & Action ID & \# Participants & Average Time Spent (mm:ss)\\
		\midrule 
		Reading code cell & A1 & 14 & 06:46 \\
		Reading markdown & A2 & 14 & 05:51 \\
		Writing/Editing code & A3 & 14 & 03:21 \\
		Using a search engine & A4 & 14 & 01:55 \\
		Looking at documentation & A5 & 13 & 03:00 \\
		Checks code output & A6 & 13 & 02:24 \\
		Inspecting DataFrame & A7 & 12 & 04:07 \\
		Looking at an online resource & A8 & 12 & 03:06 \\
		Inspecting graph & A9 & 11 & 01:53 \\
		Inspecting CSV file & A10 & 5 & 02:40 \\
		Reading an error message & A11 & 3 & 01:51 \\
		Adding a comment & A12 & 2 & 08:42 \\
		\bottomrule
	\end{tabular}
\end{table*}

Our participants performed various actions while analyzing the notebooks to find errors. 
In this context, an action is an (atomic) activity such as reading a markdown/code cell or examining a CSV file. 
Table \ref{tbl:timeaction} lists the number of participants who performed each action and the average amount of time all participants spent per action. 
There were some actions that participants always used, such as reading code and markdown cells, writing or editing code, and using the search engine. 
Other actions often used included looking at the documentation, checking code output, and inspecting dataframes. 
Finally, a few actions were only occasionally used, such as inspecting a CSV file or adding a comment. 
We describe these actions in more detail below. 
 
\cooltitle{A1: Reading a code cell.} This action refers to when participants (P1-P14) \emph{read through a code cell to understand what it was doing.} 

\cooltitle{A2: Reading markdown cells.} This action refers to when a participant (P1-P14) \emph{read through a markdown cell to gain context into what the preceding code cell tried to accomplish.}

When analyzing a notebook to find errors, participants performed actions A1 and A2 successively. 
In this scenario, P8 emphasized, \emph{``[I read] the documentation first then [I read] the code''}. 
Additionally, P12 highlighted the value of reading the markdown aloud to better understand what was going on.

\cooltitle{A3: Writing/Editing code.} This action occurred when participants (P1-P14) \emph{wrote new code in a code cell (either one they added or one present in the notebook) or edited a code cell that was initially in the notebook}. 
We observed that participants edited code for several different reasons. 
For example, P14 stated that they edited code cells to make them more readable. 
Other participants, such as P10, edited the parameters of functions to view more of the data returned by that function: for example, P10 edited calls to Pandas \mintinline{Python}{Series.nlargest()} function. 
Some participants also wrote new code into the notebooks, which served various purposes. 
For instance, P13 wrote code during their analysis of the nba\_analysis\_A notebook to verify if two sets of rows in the \mintinline{Python}{nba} dataframe were the same. 
Both P6 and P11 wrote code to perform type checking through the use of Python's \mintinline{Python}{type()} method.

\cooltitle{A4: Using a search engine. } All participants used a search engine to \emph{access some online resource or documentation page}. 
Typically participants transitioned from the notebook to the search engine and then to either an online resource or a documentation page. 
Depending on whether or not the initial search result was helpful, they would return to the notebook or select another result from the search engine. 
The \textbf{Search Engine} action is highly associated with both \textbf{A5: Looking at documentation} and \textbf{A8: Looking at an online resource}. 
We define online resources as any website other than a documentation page. 
Table \ref{tbl:onlineresources} shows the most commonly accessed documentation websites and online resources.

\begin{table}[!h]
    \small
    \centering
    \caption{Number of Visits. \textdagger~ indicates a Documentation page. The remainder are Online Resources. Fourteen other Online Resources were each visited between one and three times.}
    \label{tbl:onlineresources}
    \begin{tabular}{lc}
        \toprule
        Resource         & Number of Visits \\
        \midrule
        Pandas     \textdagger           & 58               \\
        Plotnine    \textdagger          & 28               \\
        Statsmodels  \textdagger       & 21               \\
        stackoverflow.com       & 14 \\
 		geeksforgeeks.org       & 6  \\
        investopedia.com        & 6  \\
		Numpy      \textdagger           & 4                \\
        Scipy.stats     \textdagger      & 4                \\
            tutorialspoint.com      & 4  \\
        w3schools.com           & 4  \\
        \bottomrule
    \end{tabular}
\end{table}
 
\cooltitle{A6: Checking code output.} Commonly, participants (P1-P12, P14) \emph{inspected the output of a code cell visually, either one initially present in the notebook or one which the participant added}.

\cooltitle{A7: Inspecting dataframe.} We observed that participants (P1-P4, P6-P12, P14) used the \mintinline[breaklines, breakbefore=.]{Python}{Dataframe.head()} method to \emph{visually inspect the dataframe, either to gain a preliminary understanding of the data or to check if anything seemed out of place}.

\cooltitle{A9: Inspecting a graph.} Participants (P3, P5-P14) performed this action to \emph{visually inspect any graph present in the notebook}. 

\cooltitle{A10: Inspecting CSV File.} In a similar situation to A7, participants (P5, P6, P9, P10, P13) inspected the data in its raw state. 
For instance, P13 pointed out that when using Jupyter Notebooks, they do not use the CSV viewer native to Jupyter; instead, they use an alternative application. 
Likewise, P9 indicated they use Notepad++ to view their CSV files. 
Finally, P10 highlighted that they inspect the CSV file when they are unsure how to perform a task programmatically. 
In this matter, P10 stated, \emph{``I'm just learning Python, so I can't...list these things, I actually refer to the CSV quite a bit''}. 

\cooltitle{A11: Reading an error message.} This action occurred when participants (P1, P3, P5, P8-P11) changed the notebook as initially the notebooks did not return any error messages. 
In this scenario, P10 pointed out that when they see an error message, they \emph{``don't have a clue''}. 

\cooltitle{A12: Adding a comment.} This action occurred when participants (P5, P11, P14) \emph{added a comment to a code cell either in the form of a note or to comment out code}. 

While these actions capture the more atomic tasks our participants performed, we also observed that several actions were used together to form strategies that helped participants find or understand the cause of errors. 
In the remainder of this section, we describe these strategies in more detail.  

\subsection{Error-Finding Strategies}\label{ssec:strategies}
Participants performed many of the preceding actions together to serve a particular purpose. 
We call a collection of related actions a \textit{strategy}. 
We describe the strategies we found in detail. Table \ref{tbl:strategytable} gives a brief description along with the number of participants who used each strategy.

\begin{table*}
    \small
    \centering
    \caption{Strategy Descriptions}
    \label{tbl:strategytable}
    \begin{tabular}{lp{0.45\linewidth}cc}
     \toprule
     Strategy &  Description & \# Participants & Associated Actions \\
     \midrule 
Search Engine-Driven Approach & Using the search engine and external resources to gather useful information.                     & 14              & A4, A5, A8          \\
Assume and (Sometimes) Check  & Making an assumption related to the notebook or to an API call and sometimes checking it.                                                   & 14              & A3, A7, A12         \\
Expectation Confirmation      & The participant's expectation, set up by an explanatory markdown cell, of what a code cell does cannot be confirmed upon seeing its output. & 7               & A1, A2              \\
Once-Over                     & Briefly browsing through the notebook in order to gain a preliminary understanding of what it contains.                                     & 4               & A1, A2, A6          \\
Re-implement to Check         & Re-implementing a code cell using a different syntax in order to check its validity.                                                        & 3               & A3, A6              \\
Key Information               & Extracting need-to-know information from a markdown cell and placing it in a comment inside the related code cell.                 & 1               & A2, A10             \\
Start With What You Know      & Starting at a point in the notebook which is most familiar.                                                                                 & 1               & A1, A2              \\
     \bottomrule
    \end{tabular}
\end{table*}

\cooltitle{Search Engine-Driven Approach.} 
The most common strategy we observed was the \emph{search engine-driven approach}, which every participant used. 
All participants made several transitions from the notebook to the search engine, then to an external resource, until they found a helpful online resource or documentation page.

Participants outlined three different reasons for using the search engine and external resources. 
First, they used the search engine as a first step to gather a solution from an online reference. For instance, P12 highlighted that they use Google quite often when using a Jupyter Notebook, and without it, they would not know what to do. Not knowing what to do without the search engine hints at being dependent on it; it is unknown whether this is caused by a lack of general programming knowledge or knowledge of a specific API, such as Pandas.

Second, the search engine was also used as a confirmatory aid; this happened when participants had prior knowledge. 
However, they sought supplementary expertise to confirm or refresh their intuition. 
For instance, P7 stated they often remember general concepts but use the search engine to gather information about what some specific terms mean to interpret them correctly, such as when P7 gathered information about interpreting the results of an ordinary least squares (OLS) regression.

Finally, participants also used the search engine to gather code snippets as potential solutions. 
P8 emphasized that their particular use of the search engine was to find code snippets that could help them fix the errors they identified. 

\cooltitle{Assume and (Sometimes) Check.}  
Participants would only cursorily inspect a code cell, see what the code is doing, and return to it only when they identified a potential problem in their theory of the notebook's execution. 
They then made an assumption about where in the preceding cells that problem happened,  
and then examined that code in more detail than they did on their first pass over it. 
However, participants ``sometimes'' left some assumptions unchecked. 
This may be due to the contrived nature of the study (fixing the bug was not part of the task). 
When participants did check assumptions, they wrote new code in the notebook or examined the dataframe/CSV file. 

Consider the error and thought processes of P8 
while they use the \emph{assume and (sometimes) check} strategy to determine the error described in Listing \ref{lst:dataerror} (code cell 3 of the notebook NBA\_Analysis\_A ). 
P8 began analyzing the notebook using a \emph{once-over} (see the next strategy) and noticed in a later code cell that the given mean of the height column was roughly 12. 
They then remarked that a \emph{``mean height of 12 doesn't seem to make a lot of sense''} (since height in cm should be (broadly) greater than 100cm and less than 225cm).
They then transitioned to read code cell 3 (Listing \ref{lst:dataerror} line 2), which cleaned and adjusted the height column. 
Rereading the code cell led them to \textit{assume} something must have gone wrong in that notebook cell. 
They then inspected the original dataframe and made another assumption: \emph{``Here the measurements are presumably in feet-inches and over here we have them in cm''}. 
This second assumption is an example of assuming the purpose of a series of method calls.
A closer inspection of code cell 3 allowed them to identify the error as replacing inches with the empty string and not accurately converting feet-inches to cm.

\cooltitle{Expectation Confirmation.} Seven participants (P1, P3, P5, P7, P10, P11, P13) indicated that a discrepancy between explanatory text in a markdown cell and the subsequent code cell helped them identify an error. 
P7 described the explanatory markdown as a \emph{``guidance for what I should be looking for''}, and that when a difference occurred between the markdown and the code cell, they knew something was incorrect. 
Additionally, P5 used an analogy to describe the discrepancy between the markdown and code, stating, \emph{``It's basically like `Hey, we did this' and then [I] look at the code and it's like `No, you didn't.'''} Finally, P11 emphasized, \emph{``what I was expecting is that we want a percentage and this is obviously not a percentage''}, outlining how the markdown sets their expectations. 
When the code does not fulfill these expectations, they know something is wrong.

\cooltitle{Once-Over.} Four participants (P2, P6, P8, P14) used this strategy, which involves looking through the notebook to gain a preliminary understanding. 
This strategy consists of reading markdown and code cells, running code cells and briefly checking their output, and generally inspecting the notebook's initial state. 
A once-over gives a basic understanding of what the notebook is doing without too much detail. 
All four of the participants, when using the \emph{once-over} strategy, employed different language to describe it. 
For example, P2 stated they were getting \emph{``a lay of the land''}.

\cooltitle{Re-implement to Check.} The \emph{re-implement to check} strategy was used by three participants (P1, P6, P11) and implies rewriting a code cell using a different syntax and then comparing the results of both to see if there are any differences. 
For example, P1 wrongly believed that Listing \ref{lst:P1initial} was incorrect due to the \mintinline{Python}{.agg()} syntax.
They continued to add a new cell and rewrite the code (Listing \ref{P1reimplemented}), only to find that they produced the same result.

\begin{listing}
    \centering
    \begin{minted}[fontsize=\small, frame=single, numbersep=-10pt, linenos, breaklines]{Python}
    nba[(nba['Position'] == 'PG') | (nba['Position'] == 'SG')].groupby('Position').agg(Height=('Height', 'mean'))
    \end{minted}
    \caption{Code snippet P1 wrongly thought was incorrect.}
    \label{lst:P1initial}
\end{listing}

\begin{listing}
    \centering
    \begin{minted}[fontsize=\small, frame=single, xleftmargin=\parindent, numbersep=-10pt, linenos, breaklines]{Python}
    nba[(nba['Position'] == 'PG') | (nba['Position'] == 'SG')].groupby('Position').agg('Height').mean()
    \end{minted}
    \caption{P1's re-implementation of Listing~\ref{lst:P1initial}.}
    \label{P1reimplemented}
\end{listing}

\begin{table*}[!ht]
    \small
    \centering
    \caption{Relating Strategies (from Table \ref{tbl:strategytable}) and Error Type. A dash (-) indicates no use. The once-over strategy was not used for any of the error types. There are a maximum of 21 programming errors, 12 statistical errors, and 7 data errors. }
    \label{tbl:relatingstrats}
    \begin{tabular}{lcccc} 
        \toprule
        \textbf{Strategy} & \textbf{Error Type }& \textbf{Times Used} & \textbf{Errors Found} & \textbf{Percentage}  \\ \midrule
        \multirow{3}{*}{\textbf{Search Engine-Driven Approach}} & Programming  & \textbf{26} & 11 & 52.4\% \\
                                                                & Statistical  & \textbf{16} & \textbf{7} & 53.9\% \\
                                                                & Data         & \textbf{10 }& 2 & 28.6\% \\\midrule
        \multirow{3}{*}{\textbf{Assume and Check}}  & Programming  & 13 & 7 & 33.3\% \\
                                                    & Statistical  & 8  & 4 & 30.8\%\\
                                                    & Data         & 5  & 4 & 57.1\% \\\midrule
        \multirow{3}{*}{\textbf{Expectation Confirmation}} & Programming & 20 & \textbf{17} & 81.0\% \\
                                                        & Statistical & 6  & 0  & 0\% \\
                                                        & Data        & 7  & \textbf{7}  & 100\% \\\midrule
        \multirow{3}{*}{\textbf{Re-implement to Check}} & Programming  & 1 & 0 & 0\%  \\
                                                        & Statistical  & - & - & -    \\
                                                        & Data         & 1 & 0 & 0\%  \\\midrule
        \multirow{3}{*}{\textbf{Start With What You Know}} & Programming & 1 & 0 & 0\% \\
                                                           & Statistical & - & - & -\% \\
                                                           & Data        & 1 & 0 & 0\% \\\midrule
        \multirow{3}{*}{\textbf{Key Information}} & Programming & 3  & 2 & 9.52\% \\
                                                  & Statistical & 2  & 1 & 7.69\% \\
                                                  & Data        & -  & - & -\%\\
        \bottomrule
    \end{tabular}
\end{table*}

P6 stated that they would have shown a correlation by plotting rather than using an OLS regression, but they did not re-implement this code cell as they were unfamiliar with the Plotnine package used to generate the plots. 
While not precisely re-implementation, P11 wrote pseudocode before looking at a code cell and after reading its markdown explanation. 
They then compared this pseudocode to the actual code, and if similar, P11 believed this code cell was correct and continued to a new cell. 
Additionally, participants combined this pseudocode strategy with a re-implementation to further validate a given code snippet.

\cooltitle{Key Information.} The \emph{Key Information} strategy was used four times by P5 and describes extracting only the information you need from the markdown description of a code cell; P5 then placed this information inside the code cell as a comment. 
Extraction of the key information allowed P5 to get the information closer to the code, and reduced the number of times they re-read a markdown cell to remind themselves of what a code cell was doing. 
In addition, they highlighted how extracting the key information allowed for easier comparison of the code and markdown, and eliminated any extraneous information they did not need to know. 
Using the \emph{Key Information} strategy allowed P5 to more easily employ the \emph{Expectation Confirmation} strategy.

\cooltitle{Start With What You Know.} P5 employed another strategy named \emph{Start With What You Know}, which involved analyzing parts of the notebook they were familiar with first. 
They mentioned that doing so made them \emph{``feel more confident''}, and that starting with the topics they were more familiar with gave them a better chance to find errors. 
This confidence then allowed them to find errors in the other sections of the notebook as they were better able to understand the nature of the errors.

\subsection{Strategy Success}
We now describe how the strategies outlined in Section \ref{ssec:strategies} were used to find the various types of errors present in each notebook. 
As our study was exploratory, we do not make any claim that these are the best strategies for finding a particular type of error (such a claim would require future work).
Recall that our study included the analysis of three types of errors from \cite{mcnamara22}: programming errors, statistical errors, and data/context errors (see Section \ref{sec:material}). The error types are not mutually exclusive and a given error can belong to more than one error type.
While some strategies were less successful, they are still worth examining. First, we cannot claim that unsuccessful strategies might not be successful in different contexts. Second, these strategies, if repeatedly used, might become anti-patterns for debugging that are important to know about and to avoid. Finally, strategy success can be user-dependent. 
Murphy {\it et al.} \cite{Murphy2008} also found that the same strategy can be effective or ineffective, depending on the way it is used.

Table \ref{tbl:relatingstrats} outlines the number of times our participants used each strategy per error type, the number of errors found per strategy, and the percentage of total errors found by each strategy. We report on all seven strategies.
The most successful strategies are \textbf{Expectation Confirmation} and \textbf{Search Engine-Driven Approach}. The \textbf{Expectation Confirmation} strategy success is influenced by the markdown present in our notebooks. The markdown description set expectations for our participants.
When the participants read the code following the descriptive text, they contrasted their expectations of what the code was supposed to do with what the code actually did. We note that in practice, Pimentel \emph{et al.} found that notebooks contain very little markdown \cite{Pimentel2019}. 

Additionally, we note that the efficacy of the \textbf{Search Engine-Driven Approach} is associated with the popularity of using online resources to guide users of Jupyter Notebooks \cite{Koenzen2020}, as pointed out by participants P7, P8, and P12. 
Koenzen \emph{et al.} similarly determined that code reuse in Jupyter Notebooks most commonly comes from searches on the web, most often from websites that provide a tutorial, followed by API documentation \cite{Koenzen2020}. 
    
\section{Discussion}

We discuss the implications of our work to Jupyter notebook users, notebook tool designers, and educators. We also provide insights about the differences in debugging notebooks and non-notebook code, and the threats to the validity of our work.

\subsection{Implications}

In general, the error-finding strategies we identified point to the need for more tool support when developing Jupyter Notebooks to bring them to the same level as support in more mature \idecode{} tools. 
For example, the release of Jupyter Lab 3.0 introduced a visual debugger that can be used to step through code or to check the value of a variable~\cite{Tuloup2021}. 
This need for more tool support is suggested by other studies as well \cite{Chattopadhyay2020,yang21ase}.

We also uncovered two strategies that are not common in other approaches and may be specific to notebooks: 
\textit{Re-implement to Check} and \textit{Start With What You Know}.
We discuss the implications of these two strategies for notebook stakeholders below.

\cooltitle{Re-implement to Check:}
\begin{description}
	\item \textbf{Tool designers} could implement a tool which supplies the user with code snippets that use a different implementation so they could compare if the results are the same.
	\item \textbf{Users}, if unsure what a particular code cell does, could be advised to re-implement the code to increase their understanding of the code in question and make it easier to identify an error.
	\item \textbf{Educators}, when teaching students how to perform a task, could help students be aware that there may be more than one correct implementation. This would mitigate the false assumption that unfamiliar implementations (e.g., Pythonic list comprehensions) are incorrect.
\end{description}

\cooltitle{Start With What You Know:}
\begin{description}
	\item \textbf{Tool designers} could provide complexity measures for code cells so that the user can compare their own previous experience with the complexity of the cell to gauge where to start.
	\item \textbf{Users} could be advised to start by self-reflecting on their skills in code understanding (e.g., data cleaning vs. statistical analysis) and start the process of error identification in cells by leveraging that skill. This may make the process of error finding easier as the user is more familiar with this approach and can build confidence in error finding.
	\item \textbf{Educators} should understand what students are most familiar with (statistical, code, data domains) and then help them build knowledge in other areas. They could include a component on Jupyter or other notebook-specific debugging skills, such as the shift tab shortcut to access documentation.
\end{description}

\subsection{Comparing Debugging Notebooks and Debugging Non-Notebook Code}

The development of \idecode{} differs from the development of computational notebooks. 
The type of problem managed in a notebook involves more data wrangling, experimentation, and analysis code. 
Following the study which inspired our research \cite{mcnamara22}, our study separated these into potential problems with statistics, programming, and data / domain knowledge. 
The notebook environment also has a literate programming component that goes beyond code comments, with markdown cells that can be used to describe the purpose of the code. 

Furthermore, non-notebook IDEs have robust tool support for debugging, for example, setting breakpoints in IntelliJ.
However, in the traditional computational notebook interface (say Jupyter Notebook), debugging is not specifically supported by the tool. 
Data science tools are actively working to fix this, for example, Jupyter Lab's debugger \cite{Tuloup2021} and RStudio's debugging interface. IDEs are also now able to integrate notebook code into the IDE directly, such as with Visual Studio Code.

Given these differences, we ask whether error identification approaches for Jupyter Notebooks are also different. 
This study identified several strategies participants used to identify errors in Jupyter Notebooks. 
Other researchers have identified strategies for debugging \idecode{}. 
Table \ref{tbl:similarstrategies} outlines strategies identified by Murphy \emph{et al.} and Whalley \emph{et al.}, that are similar to those we have identified \cite{Murphy2008,Whalley2021}.

\begin{table}[!ht]
    \small
    \centering
    \caption{Strategies Similar to Those We Identified}
    \label{tbl:similarstrategies}
    \begin{tabularx}{\columnwidth}{lX}
    \toprule
    Strategies We Identified                & Similar Strategies \\                
    \midrule
    Search Engine-Driven Approach & Using Resources \cite{Murphy2008} \\
    Assume and (Sometimes) Check  & Information Gathering \cite{Whalley2021}, Bug Location \cite{Whalley2021} \\
    Expectation Confirmation         & Pattern Matching \cite{Murphy2008} \\
    Once-over                     & Gain Domain Knowledge \cite{Murphy2008}, Understanding the Code \cite{Murphy2008}, Static Code Comprehension \cite{Whalley2021} \\
    Re-implement to Check         & N/A \\
    Key Information               & Understanding Code \cite{Murphy2008}, Static Code Comprehension \cite{Whalley2021}\\
    Start With What You Know      & N/A \\
    \bottomrule
    \end{tabularx}
\end{table}

The \textbf{Search Engine-Driven Approach} strategy is closely related to the \textbf{Using Resources} strategy identified by Murphy \emph{et al.} in \cite{Murphy2008}. 
Both strategies involve the use of documentation and tutorials. 
The difference between these two strategies is that we observed our participants using the search engine as their gateway to many resources. 
Murphy \emph{et al.} make no mention of the search engine. 

Both the \textbf{Information Gathering} and \textbf{Bug Location} strategies identified by Whalley \emph{et al.} mention the use of speculation and guessing about the locations and causes of bugs. 
Alaboudi and LaToza also report on using hypotheses as a debugging aid \cite{Alaboudi2020}.
Our \textbf{Assume and (Sometimes) Check} strategy is similar, based on making an assumption and optionally checking that assumption. 

In their description of the \textbf{Pattern Matching} strategy, Murphy \emph{et al.} state that their participants found bugs due to things not ``looking right''. 
In our notebook study, this was made more explicit than the heuristics Murphy \emph{et al.} describe. Our participants were able to identify errors when a code cell seemed like it was not correct based on a description given in a markdown cell (the \textbf{Expectation Confirmation} strategy). 
The breakdown of an expectation could be thought of as pattern matching as our participants were attempting to match the pattern of what they were told the code was trying to accomplish to what they could observe the code doing. However, the presence of explicit documentation makes this strategy quite successful (at least for our example notebooks).

The \textbf{Once-over} and \textbf{Key Information} strategies identified by us are both similar to the \textbf{Understanding Code} and \textbf{Static Code Comprehension} strategies identified by Murphy \emph{et al.} and Whalley \emph{et al.}, respectively. 
In addition, the \textbf{Once-over} strategy is similar to the \textbf{Gain Domain Knowledge} strategy identified by Murphy \emph{et al.} 
Both of our strategies were used in order to gain understanding about the contents of the notebook, and involve comprehending both the code and the markdown, much like the \textbf{Understanding Code} and \textbf{Static Code Comprehension} strategies are about comprehending code. 
The \textbf{Once-over} strategy was used to gain domain knowledge in the sense that the four participants who used this strategy did so to gain a brief understanding of the domain the notebook covered.

One key difference between notebooks and \idecode{} is that code cells are capable of independent output, closer to a Read-Eval-Print Loop session than debugging a complete source file. 
This difference may be to blame for why the \textbf{Re-implement to Check} and \textbf{Start With What You Know} strategies were not observed in other literature related to debugging \idecode{}.
As the code in notebooks is often more granular and independent, these strategies are more viable when used in a notebook debugging context. 
This independence and high granularity allows for easier isolation of changes and re-implementations as a single unit of code in notebooks than in \idecode{}.
The value of these strategies, however, may be dependent on the user's level of experience. 
For example, the \textbf{Re-implement to Check} strategy would more likely be adopted by users who know more than one way to implement a given task. 
On the other hand, the \textbf{Start With What You Know} strategy is more likely to be used by novice users that may want to stay within their comfort zone for as long as possible.

We found that debugging \idecode{} differs from debugging computational notebooks in a few ways. 
One, the type of development is different: there are more data science-related tasks such as data wrangling. 
Two, the development tools are at different levels of maturity when it comes to debugging support.
Three, while five out of seven of the strategies we observed are related to \idecode{} debugging strategies identified in the literature, we found that two strategies were not found in \idecode{} studies. 
We also saw differences in how \textbf{Expectation Confirmation} and \textbf{Assume and (Sometimes) Check} are conducted in practice, given the way a notebook isolates individual code cells.

\section{Threats to Validity}
In the following, we address the validity of this study in the context of qualitative research~\cite{guba1981criteria,korstjens2018series}.

\cooltitle{Internal validity.}
We did not impose time-constraints on our participants, and they were assured our study was not a test of their skill. 
However, given the nature of the task, it is possible our participants felt pressure to perform well.  
Due to this pressure, participants may have overlooked errors in the Jupyter Notebooks. 
However, during the interviews we conducted immediately after the tasks, we did not detect that our participants felt any undue stress due to the study.
    
\cooltitle{Construct validity.}
Our study prompt and task description may have influenced participants to perform actions which were not part of their typical error identification process in Jupyter Notebooks.
For instance, modifying the notebook, searching documentation, and using the internet for help may not have been naturalistic behaviours. 
To mitigate this threat, we adopted multiple strategies, such as two rounds of pilots, to ensure comprehensibility and raise the realism of the tasks. 
In addition, task descriptions and scripts were reviewed and validated by a domain expert, and the task was confirmed to be within the recruited participants' skill level.

\cooltitle{External validity.}
The primary threat to external validity is how we recruited and selected participants. 
We used convenience sampling methods to recruit participants from upper-level undergraduate and graduate-level courses at the university.
Therefore, most of our participants were students who used Jupyter Notebooks for school assignments and not professionally. 
However, we designed the tasks according to our participants' skill levels, and the context of tasks was fairly approachable (elections and sports) by any participant independently of their academic background. Our participants did not express unfamiliarity with the domain.
However, some participants expressed unfamiliarity with specific packages imported into the notebook, namely Pandas and Plotnine. 
Another threat to external validity is the inclusion of markdown cells in the notebooks which may not reflect real-world notebooks \cite{Pimentel2019}.

\cooltitle{Reliability.}
The open coding process was performed by one researcher, the first author of this paper. 
To reduce potential researcher bias and subjectivity, we conducted several discussion sessions to iteratively build a codebook. 
We confirmed with the feedback of an expert reviewer, the fourth author of this paper, to raise the reliability and maturity of our findings.

\section{Conclusion}

We conducted an observational study with fourteen participants, mostly university students from varying technical backgrounds, and observed the strategies these Jupyter Notebook users employed to identify errors seeded in four sample notebooks. 
The most commonly used strategy we observed was using the search engine to find external help such as API documentation or websites that provide a tutorial. 
However, the most successful strategy was Expectation Confirmation, when they discovered a mismatch between the description and the code itself. 
We identified some implications for practice, including the need for better debugging support in notebooks, and showed that while there are similarities with \idecode{}, debugging in notebooks leverages notebook-only properties such as code cell independence and hidden state. 
Out of the seven identified strategies, five had been previously identified in the literature as debugging strategies for non-notebook code, while two are novel to the notebook environment.
Future work could involve collaborating with members of the original RMarkdown study to compare the error finding strategies of data scientists between the respective studies.
We hope our insights will help both notebook tool designers and educators improve how data scientists discover errors more easily in their notebooks.


\section*{Acknowledgments}
We thank Kelly Bodwin and Ian Flores Siaca for sharing the initial RMarkdown notebooks;
Hunter Glanz for reviewing our Python notebook translations;
and Amelia McNamara, Amal Abel-Ghani, Philipp Burckhardt, Allison Theobold and Greg Wilson for the study idea.
We acknowledge the support of the Natural Sciences and Engineering Research Council of Canada and Venture for Canada. 

\onecolumn
\begin{multicols}{2}
\bibliographystyle{ACM-Reference-Format}
\bibliography{paper}


\begin{thebibliography}{00}


\ifx \showCODEN    \undefined \def \showCODEN     #1{\unskip}     \fi
\ifx \showDOI      \undefined \def \showDOI       #1{#1}\fi
\ifx \showISBNx    \undefined \def \showISBNx     #1{\unskip}     \fi
\ifx \showISBNxiii \undefined \def \showISBNxiii  #1{\unskip}     \fi
\ifx \showISSN     \undefined \def \showISSN      #1{\unskip}     \fi
\ifx \showLCCN     \undefined \def \showLCCN      #1{\unskip}     \fi
\ifx \shownote     \undefined \def \shownote      #1{#1}          \fi
\ifx \showarticletitle \undefined \def \showarticletitle #1{#1}   \fi
\ifx \showURL      \undefined \def \showURL       {\relax}        \fi
\providecommand\bibfield[2]{#2}
\providecommand\bibinfo[2]{#2}
\providecommand\natexlab[1]{#1}
\providecommand\showeprint[2][]{arXiv:#2}

\bibitem[\protect\citeauthoryear{Alaboudi and LaToza}{Alaboudi and
  LaToza}{2020}]%
        {Alaboudi2020}
\bibfield{author}{\bibinfo{person}{Abdulaziz Alaboudi} {and}
  \bibinfo{person}{Thomas~D. LaToza}.} \bibinfo{year}{2020}\natexlab{}.
\newblock \showarticletitle{Using Hypotheses as a Debugging Aid}. In
  \bibinfo{booktitle}{{\em 2020 IEEE Symposium on Visual Languages and
  Human-Centric Computing (VL/HCC)}}. \bibinfo{publisher}{IEEE},
  \bibinfo{address}{Dunedin, New Zealand}, \bibinfo{pages}{1--9}.
\newblock
\showDOI{%
\url{https://doi.org/10.1109/VL/HCC50065.2020.9127273}}


\bibitem[\protect\citeauthoryear{Alaboudi and LaToza}{Alaboudi and
  LaToza}{2021}]%
        {Alaboudi2021}
\bibfield{author}{\bibinfo{person}{Abdulaziz Alaboudi} {and}
  \bibinfo{person}{Thomas~D. LaToza}.} \bibinfo{year}{2021}\natexlab{}.
\newblock \showarticletitle{An Exploratory Study of Debugging Episodes}.
\newblock \bibinfo{journal}{{\em CoRR\/}}  \bibinfo{volume}{abs/2105.02162}
  (\bibinfo{year}{2021}).
\newblock
\showeprint{2105.02162}
\showURL{%
\url{https://arxiv.org/abs/2105.02162}}


\bibitem[\protect\citeauthoryear{Bodwin, Flores~Siaca, and Robinson}{Bodwin
  et~al\mbox{.}}{2022}]%
        {bodwin22zenodo}
\bibfield{author}{\bibinfo{person}{Kelly~Nicole Bodwin}, \bibinfo{person}{Ian
  Flores~Siaca}, {and} \bibinfo{person}{Derek Robinson}.}
  \bibinfo{year}{2022}\natexlab{}.
\newblock \bibinfo{title}{{Materials for "Looks okay to me": A study of best
  practice in data analysis code review.}}
\newblock   (\bibinfo{date}{April} \bibinfo{year}{2022}).
\newblock
\showDOI{%
\url{https://doi.org/10.5281/zenodo.6419727}}


\bibitem[\protect\citeauthoryear{Brown, Kaiser, and Allison}{Brown
  et~al\mbox{.}}{2018}]%
        {Brown2018}
\bibfield{author}{\bibinfo{person}{Andrew~W Brown}, \bibinfo{person}{Kathryn~A
  Kaiser}, {and} \bibinfo{person}{David~B Allison}.}
  \bibinfo{year}{2018}\natexlab{}.
\newblock \showarticletitle{Issues with data and analyses: Errors, underlying
  themes, and potential solutions}.
\newblock \bibinfo{journal}{{\em Proceedings of the National Academy of
  Sciences\/}} \bibinfo{volume}{115}, \bibinfo{number}{11}
  (\bibinfo{year}{2018}), \bibinfo{pages}{2563--2570}.
\newblock


\bibitem[\protect\citeauthoryear{Cardoso, Leit{\~a}o, and Teixeira}{Cardoso
  et~al\mbox{.}}{2019}]%
        {Cardoso2018}
\bibfield{author}{\bibinfo{person}{Alberto Cardoso}, \bibinfo{person}{Joaquim
  Leit{\~a}o}, {and} \bibinfo{person}{C{\'e}sar Teixeira}.}
  \bibinfo{year}{2019}\natexlab{}.
\newblock \showarticletitle{Using the Jupyter Notebook as a Tool to Support the
  Teaching and Learning Processes in Engineering Courses}. In
  \bibinfo{booktitle}{{\em The Challenges of the Digital Transformation in
  Education}}, \bibfield{editor}{\bibinfo{person}{Michael~E. Auer} {and}
  \bibinfo{person}{Thrasyvoulos Tsiatsos}} (Eds.). \bibinfo{publisher}{Springer
  International Publishing}, \bibinfo{address}{Cham},
  \bibinfo{pages}{227--236}.
\newblock
\showISBNx{978-3-030-11935-5}


\bibitem[\protect\citeauthoryear{Chattopadhyay, Prasad, Henley, Sarma, and
  Barik}{Chattopadhyay et~al\mbox{.}}{2020}]%
        {Chattopadhyay2020}
\bibfield{author}{\bibinfo{person}{Souti Chattopadhyay},
  \bibinfo{person}{Ishita Prasad}, \bibinfo{person}{Austin~Z. Henley},
  \bibinfo{person}{Anita Sarma}, {and} \bibinfo{person}{Titus Barik}.}
  \bibinfo{year}{2020}\natexlab{}.
\newblock \showarticletitle{What's Wrong with Computational Notebooks? Pain
  Points, Needs, and Design Opportunities}. In \bibinfo{booktitle}{{\em
  Proceedings of the 2020 CHI Conference on Human Factors in Computing
  Systems}}. \bibinfo{publisher}{Association for Computing Machinery},
  \bibinfo{address}{New York, NY, USA}, \bibinfo{pages}{1–12}.
\newblock
\showISBNx{9781450367080}
\showURL{%
\url{https://doi.org/10.1145/3313831.3376729}}


\bibitem[\protect\citeauthoryear{Dragicevic, Jansen, Sarma, Kay, and
  Chevalier}{Dragicevic et~al\mbox{.}}{2019}]%
        {Dragicevic2019}
\bibfield{author}{\bibinfo{person}{Pierre Dragicevic}, \bibinfo{person}{Yvonne
  Jansen}, \bibinfo{person}{Abhraneel Sarma}, \bibinfo{person}{Matthew Kay},
  {and} \bibinfo{person}{Fanny Chevalier}.} \bibinfo{year}{2019}\natexlab{}.
\newblock \showarticletitle{Increasing the Transparency of Research Papers with
  Explorable Multiverse Analyses}. In \bibinfo{booktitle}{{\em Proceedings of
  the 2019 {CHI} Conference on Human Factors in Computing Systems}}.
\newblock
\showDOI{%
\url{https://doi.org/10.1145/3290605.3300295}}


\bibitem[\protect\citeauthoryear{Guba}{Guba}{1981}]%
        {guba1981criteria}
\bibfield{author}{\bibinfo{person}{Egon~G Guba}.}
  \bibinfo{year}{1981}\natexlab{}.
\newblock \showarticletitle{Criteria for assessing the trustworthiness of
  naturalistic inquiries}.
\newblock \bibinfo{journal}{{\em Educational Technology research and
  development\/}} \bibinfo{volume}{29}, \bibinfo{number}{2}
  (\bibinfo{year}{1981}), \bibinfo{pages}{75--91}.
\newblock


\bibitem[\protect\citeauthoryear{Head, Hohman, Barik, Drucker, and DeLine}{Head
  et~al\mbox{.}}{2019}]%
        {Head2019}
\bibfield{author}{\bibinfo{person}{Andrew Head}, \bibinfo{person}{Fred Hohman},
  \bibinfo{person}{Titus Barik}, \bibinfo{person}{Steven~M Drucker}, {and}
  \bibinfo{person}{Robert DeLine}.} \bibinfo{year}{2019}\natexlab{}.
\newblock \showarticletitle{Managing messes in computational notebooks}. In
  \bibinfo{booktitle}{{\em Proceedings of the 2019 CHI Conference on Human
  Factors in Computing Systems}}. \bibinfo{pages}{1--12}.
\newblock


\bibitem[\protect\citeauthoryear{Kery, Radensky, Arya, John, and Myers}{Kery
  et~al\mbox{.}}{2018}]%
        {Kery2018}
\bibfield{author}{\bibinfo{person}{Mary~Beth Kery}, \bibinfo{person}{Marissa
  Radensky}, \bibinfo{person}{Mahima Arya}, \bibinfo{person}{Bonnie~E John},
  {and} \bibinfo{person}{Brad~A Myers}.} \bibinfo{year}{2018}\natexlab{}.
\newblock \showarticletitle{The {{Story}} in the {{Notebook}}: {{Exploratory
  Data Science Using}} a {{Literate Programming Tool}}}. In
  \bibinfo{booktitle}{{\em Proceedings of the 2018 {{CHI Conference}} on
  {{Human Factors}} in {{Computing Systems}}}} {\em (\bibinfo{series}{{{CHI}}
  '18})}. \bibinfo{publisher}{{Association for Computing Machinery}},
  \bibinfo{address}{{New York, NY, USA}}.
\newblock
\showISBNx{978-1-4503-5620-6}
\showDOI{%
\url{https://doi.org/10.1145/3173574.3173748}}


\bibitem[\protect\citeauthoryear{Koenzen, Ernst, and Storey}{Koenzen
  et~al\mbox{.}}{2020}]%
        {Koenzen2020}
\bibfield{author}{\bibinfo{person}{Andreas~P. Koenzen},
  \bibinfo{person}{Neil~A. Ernst}, {and} \bibinfo{person}{Margaret-Anne~D.
  Storey}.} \bibinfo{year}{2020}\natexlab{}.
\newblock \showarticletitle{Code Duplication and Reuse in Jupyter Notebooks}.
  In \bibinfo{booktitle}{{\em 2020 IEEE Symposium on Visual Languages and
  Human-Centric Computing (VL/HCC)}}. \bibinfo{publisher}{IEEE},
  \bibinfo{address}{Dunedin, New Zealand}, \bibinfo{pages}{1--9}.
\newblock
\showDOI{%
\url{https://doi.org/10.1109/VL/HCC50065.2020.9127202}}


\bibitem[\protect\citeauthoryear{Korstjens and Moser}{Korstjens and
  Moser}{2018}]%
        {korstjens2018series}
\bibfield{author}{\bibinfo{person}{Irene Korstjens} {and}
  \bibinfo{person}{Albine Moser}.} \bibinfo{year}{2018}\natexlab{}.
\newblock \showarticletitle{Series: Practical guidance to qualitative research.
  Part 4: Trustworthiness and publishing}.
\newblock \bibinfo{journal}{{\em European Journal of General Practice\/}}
  \bibinfo{volume}{24}, \bibinfo{number}{1} (\bibinfo{year}{2018}),
  \bibinfo{pages}{120--124}.
\newblock


\bibitem[\protect\citeauthoryear{McNamara, Abel-Ghani, Siaca~Flores, Bodwin,
  Theobold, Burkhardt, and Wilson}{McNamara et~al\mbox{.}}{2022}]%
        {mcnamara22}
\bibfield{author}{\bibinfo{person}{Amelia McNamara}, \bibinfo{person}{Amal
  Abel-Ghani}, \bibinfo{person}{Ian Siaca~Flores}, \bibinfo{person}{Kelly
  Bodwin}, \bibinfo{person}{Allison Theobold}, \bibinfo{person}{Philipp
  Burkhardt}, {and} \bibinfo{person}{Greg Wilson}.}
  \bibinfo{year}{2022}\natexlab{}.
\newblock \bibinfo{title}{Looks okay to me: A study of best practice in data
  analysis code review}.
\newblock   (\bibinfo{year}{2022}).
\newblock
\newblock
\shownote{in preparation.}


\bibitem[\protect\citeauthoryear{Murphy, Lewandowski, McCauley, Simon, Thomas,
  and Zander}{Murphy et~al\mbox{.}}{2008}]%
        {Murphy2008}
\bibfield{author}{\bibinfo{person}{Laurie Murphy}, \bibinfo{person}{Gary
  Lewandowski}, \bibinfo{person}{Ren{\'e}e McCauley}, \bibinfo{person}{Beth
  Simon}, \bibinfo{person}{Lynda Thomas}, {and} \bibinfo{person}{Carol
  Zander}.} \bibinfo{year}{2008}\natexlab{}.
\newblock \showarticletitle{Debugging: the good, the bad, and the quirky--a
  qualitative analysis of novices' strategies}.
\newblock \bibinfo{journal}{{\em ACM SIGCSE Bulletin\/}} \bibinfo{volume}{40},
  \bibinfo{number}{1} (\bibinfo{year}{2008}), \bibinfo{pages}{163--167}.
\newblock


\bibitem[\protect\citeauthoryear{Parente}{Parente}{2014}]%
        {Parente2014}
\bibfield{author}{\bibinfo{person}{Peter Parente}.}
  \bibinfo{year}{2014}\natexlab{}.
\newblock \bibinfo{title}{Estimate of public {J}upyter notebooks on {GitHub}}.
\newblock   (\bibinfo{year}{2014}).
\newblock
\showURL{%
\url{https://github.com/parente/nbestimate}}
\newblock
\shownote{(Accessed on 01/16/2022).}


\bibitem[\protect\citeauthoryear{Perkel}{Perkel}{2018}]%
        {Perkel2018}
\bibfield{author}{\bibinfo{person}{Jeffrey~M Perkel}.}
  \bibinfo{year}{2018}\natexlab{}.
\newblock \showarticletitle{Why Jupyter is data scientists' computational
  notebook of choice}.
\newblock \bibinfo{journal}{{\em Nature\/}} \bibinfo{volume}{563},
  \bibinfo{number}{7732} (\bibinfo{year}{2018}), \bibinfo{pages}{145--147}.
\newblock
\showDOI{%
\url{https://doi.org/10.1038/d41586-018-07196-1}}


\bibitem[\protect\citeauthoryear{Pimentel, Murta, Braganholo, and
  Freire}{Pimentel et~al\mbox{.}}{2019}]%
        {Pimentel2019}
\bibfield{author}{\bibinfo{person}{João~Felipe Pimentel},
  \bibinfo{person}{Leonardo Murta}, \bibinfo{person}{Vanessa Braganholo}, {and}
  \bibinfo{person}{Juliana Freire}.} \bibinfo{year}{2019}\natexlab{}.
\newblock \showarticletitle{A Large-Scale Study About Quality and
  Reproducibility of Jupyter Notebooks}. In \bibinfo{booktitle}{{\em 2019
  IEEE/ACM 16th International Conference on Mining Software Repositories
  (MSR)}}. \bibinfo{publisher}{IEEE}, \bibinfo{address}{Montreal, QC, Canada},
  \bibinfo{pages}{507--517}.
\newblock
\showDOI{%
\url{https://doi.org/10.1109/MSR.2019.00077}}


\bibitem[\protect\citeauthoryear{Rezig, Brahmaroutu, Tatbul, Ouzzani, Tang,
  Mattson, Madden, and Stonebraker}{Rezig et~al\mbox{.}}{2020}]%
        {Rezig2020}
\bibfield{author}{\bibinfo{person}{El~Kindi Rezig}, \bibinfo{person}{Ashrita
  Brahmaroutu}, \bibinfo{person}{Nesime Tatbul}, \bibinfo{person}{Mourad
  Ouzzani}, \bibinfo{person}{Nan Tang}, \bibinfo{person}{Timothy Mattson},
  \bibinfo{person}{Samuel Madden}, {and} \bibinfo{person}{Michael
  Stonebraker}.} \bibinfo{year}{2020}\natexlab{}.
\newblock \showarticletitle{Debugging large-scale data science pipelines using
  {D}agger}.
\newblock \bibinfo{journal}{{\em Proceedings of the {VLDB} Endowment\/}}
  \bibinfo{volume}{13}, \bibinfo{number}{12} (\bibinfo{date}{Aug.}
  \bibinfo{year}{2020}), \bibinfo{pages}{2993--2996}.
\newblock
\showDOI{%
\url{https://doi.org/10.14778/3415478.3415527}}


\bibitem[\protect\citeauthoryear{Rule, Tabard, and Hollan}{Rule
  et~al\mbox{.}}{2018}]%
        {Rule2018}
\bibfield{author}{\bibinfo{person}{Adam Rule}, \bibinfo{person}{Aur\'{e}lien
  Tabard}, {and} \bibinfo{person}{James~D. Hollan}.}
  \bibinfo{year}{2018}\natexlab{}.
\newblock \showarticletitle{Exploration and Explanation in Computational
  Notebooks}. In \bibinfo{booktitle}{{\em Proceedings of the 2018 CHI
  Conference on Human Factors in Computing Systems}}. \bibinfo{pages}{1--12}.
\newblock
\showISBNx{9781450356206}
\showURL{%
\url{https://doi.org/10.1145/3173574.3173606}}


\bibitem[\protect\citeauthoryear{Smith}{Smith}{2016}]%
        {Smith2016}
\bibfield{author}{\bibinfo{person}{Adam~A Smith}.}
  \bibinfo{year}{2016}\natexlab{}.
\newblock \showarticletitle{Teaching computer science to biologists and
  chemists, using jupyter notebooks: tutorial presentation}.
\newblock \bibinfo{journal}{{\em Journal of Computing Sciences in Colleges\/}}
  \bibinfo{volume}{32}, \bibinfo{number}{1} (\bibinfo{year}{2016}),
  \bibinfo{pages}{126--128}.
\newblock
\showDOI{%
\url{https://doi.org/10.5555/3007225.3007252}}


\bibitem[\protect\citeauthoryear{Tuloup}{Tuloup}{2021}]%
        {Tuloup2021}
\bibfield{author}{\bibinfo{person}{Jeremy Tuloup}.}
  \bibinfo{year}{2021}\natexlab{}.
\newblock \bibinfo{title}{JupyterLab 3.0 is released!. The 3.0 release of
  JupyterLab brings \ldots{}}.
\newblock   (\bibinfo{year}{2021}).
\newblock
\showURL{%
\url{https://blog.jupyter.org/jupyterlab-3-0-is-out-4f58385e25bb}}
\newblock
\shownote{(Accessed on 07/28/2021).}


\bibitem[\protect\citeauthoryear{Wang, Kuo, Li, and Zeller}{Wang
  et~al\mbox{.}}{2020a}]%
        {Wang2020Restoring}
\bibfield{author}{\bibinfo{person}{Jiawei Wang}, \bibinfo{person}{Tzu-yang
  Kuo}, \bibinfo{person}{Li Li}, {and} \bibinfo{person}{Andreas Zeller}.}
  \bibinfo{year}{2020}\natexlab{a}.
\newblock \bibinfo{booktitle}{{\em Restoring Reproducibility of Jupyter
  Notebooks}}.
\newblock \bibinfo{publisher}{Association for Computing Machinery},
  \bibinfo{address}{New York, NY, USA}, \bibinfo{pages}{288--289}.
\newblock
\showISBNx{9781450371223}
\showURL{%
\url{https://doi.org/10.1145/3377812.3390803}}


\bibitem[\protect\citeauthoryear{Wang, Li, and Zeller}{Wang
  et~al\mbox{.}}{2020b}]%
        {Wang2020Better}
\bibfield{author}{\bibinfo{person}{Jiawei Wang}, \bibinfo{person}{Li Li}, {and}
  \bibinfo{person}{Andreas Zeller}.} \bibinfo{year}{2020}\natexlab{b}.
\newblock \showarticletitle{Better Code, Better Sharing: On the Need of
  Analyzing Jupyter Notebooks}. In \bibinfo{booktitle}{{\em Proceedings of the
  ACM/IEEE 42nd International Conference on Software Engineering: New Ideas and
  Emerging Results}} {\em (\bibinfo{series}{ICSE-NIER '20})}.
  \bibinfo{publisher}{Association for Computing Machinery},
  \bibinfo{address}{New York, NY, USA}, \bibinfo{pages}{53--56}.
\newblock
\showISBNx{9781450371261}
\showDOI{%
\url{https://doi.org/10.1145/3377816.3381724}}


\bibitem[\protect\citeauthoryear{Weinman, Drucker, Barik, and DeLine}{Weinman
  et~al\mbox{.}}{2021}]%
        {Weinman2021}
\bibfield{author}{\bibinfo{person}{Nathaniel Weinman},
  \bibinfo{person}{Steven~M. Drucker}, \bibinfo{person}{Titus Barik}, {and}
  \bibinfo{person}{Robert DeLine}.} \bibinfo{year}{2021}\natexlab{}.
\newblock \showarticletitle{Fork It: Supporting Stateful Alternatives in
  Computational Notebooks}. In \bibinfo{booktitle}{{\em Proceedings of the 2021
  {CHI} Conference on Human Factors in Computing Systems}}.
\newblock
\showDOI{%
\url{https://doi.org/10.1145/3411764.3445527}}


\bibitem[\protect\citeauthoryear{Weiss}{Weiss}{2021}]%
        {Weiss2020}
\bibfield{author}{\bibinfo{person}{Charles~J. Weiss}.}
  \bibinfo{year}{2021}\natexlab{}.
\newblock \showarticletitle{A Creative Commons Textbook for Teaching Scientific
  Computing to Chemistry Students with Python and Jupyter Notebooks}.
\newblock \bibinfo{journal}{{\em Journal of Chemical Education\/}}
  \bibinfo{volume}{98}, \bibinfo{number}{2} (\bibinfo{year}{2021}),
  \bibinfo{pages}{489--494}.
\newblock
\showDOI{%
\url{https://doi.org/10.1021/acs.jchemed.0c01071}}


\bibitem[\protect\citeauthoryear{Whalley, Settle, and Luxton-Reilly}{Whalley
  et~al\mbox{.}}{2021}]%
        {Whalley2021}
\bibfield{author}{\bibinfo{person}{Jacqueline Whalley}, \bibinfo{person}{Amber
  Settle}, {and} \bibinfo{person}{Andrew Luxton-Reilly}.}
  \bibinfo{year}{2021}\natexlab{}.
\newblock \bibinfo{booktitle}{{\em Novice Reflections on Debugging}}.
\newblock \bibinfo{publisher}{Association for Computing Machinery},
  \bibinfo{address}{New York, NY, USA}, \bibinfo{pages}{73--79}.
\newblock
\showISBNx{9781450380621}
\showURL{%
\url{https://doi.org/10.1145/3408877.3432374}}


\bibitem[\protect\citeauthoryear{Wongsuphasawat, Liu, and Heer}{Wongsuphasawat
  et~al\mbox{.}}{2019}]%
        {wong19}
\bibfield{author}{\bibinfo{person}{Kanit Wongsuphasawat}, \bibinfo{person}{Yang
  Liu}, {and} \bibinfo{person}{Jeffrey Heer}.} \bibinfo{year}{2019}\natexlab{}.
\newblock \showarticletitle{Goals, Process, and Challenges of Exploratory Data
  Analysis: An Interview Study}.
\newblock \bibinfo{journal}{{\em CoRR\/}}  \bibinfo{volume}{abs/1911.00568}
  (\bibinfo{year}{2019}).
\newblock
\showeprint{1911.00568}
\showURL{%
\url{http://arxiv.org/abs/1911.00568}}


\bibitem[\protect\citeauthoryear{Yang, Zhou, Guo, and K{\"a}stner}{Yang
  et~al\mbox{.}}{2021}]%
        {yang21ase}
\bibfield{author}{\bibinfo{person}{Chenyang Yang}, \bibinfo{person}{Shurui
  Zhou}, \bibinfo{person}{Jin~LC Guo}, {and} \bibinfo{person}{Christian
  K{\"a}stner}.} \bibinfo{year}{2021}\natexlab{}.
\newblock \showarticletitle{Subtle bugs everywhere: Generating documentation
  for data wrangling code}. In \bibinfo{booktitle}{{\em Proceedings of the 36th
  IEEE/ACM International Conference on Automated Software Engineering (ASE).
  IEEE}}, Vol.~\bibinfo{volume}{11}.
\newblock


\bibitem[\protect\citeauthoryear{Zeller}{Zeller}{2009}]%
        {Zeller2009programs}
\bibfield{author}{\bibinfo{person}{Andreas Zeller}.}
  \bibinfo{year}{2009}\natexlab{}.
\newblock \showarticletitle{CHAPTER 1 - How Failures Come to Be}.
\newblock In \bibinfo{booktitle}{{\em Why Programs Fail (Second Edition)}
  (\bibinfo{edition}{second edition} ed.)},
  \bibfield{editor}{\bibinfo{person}{Andreas Zeller}} (Ed.).
  \bibinfo{publisher}{Morgan Kaufmann}, \bibinfo{address}{Boston},
  \bibinfo{pages}{1--23}.
\newblock
\showISBNx{978-0-12-374515-6}
\showDOI{%
\url{https://doi.org/10.1016/B978-0-12-374515-6.00001-0}}


\end{thebibliography}
\end{multicols}
\end{document}